\documentclass[aps,prb,twocolumn,superscriptaddress]{revtex4}
\usepackage{amsmath,amssymb}
\usepackage{bm}
\usepackage{graphicx}
\usepackage{epstopdf}
\usepackage{latexsym}
\usepackage{subfigure}
\usepackage{color}

\newcommand{\be}{\begin{equation}}
\newcommand{\ee}{\end{equation}}
\newcommand{\bea}{\begin{eqnarray}}
\newcommand{\eea}{\end{eqnarray}}

\newcommand{\eg}{{e$_g$}}

\begin{document}
\title{Landau theory of charge and spin ordering in the nickelates}
\author{SungBin Lee}
\author{Ru Chen}
\affiliation{Physics Department, University of California, Santa
  Barbara, CA, 93106}
\author{Leon Balents}
\affiliation{Kavli Institute for Theoretical Physics, University of
  California, Santa Barbara, CA, 93106-9530}

\date{\today}
\begin{abstract}
  Guided by experiment and band structure, we introduce and study a
  phenomenological Landau theory for the unusual charge and spin
  ordering associated with the Mott transition in the perovskite
  nickelates, with chemical formula RNiO$_3$, where R=Pr, Nd,Sm, Eu, Ho,
  Y, and Lu.  While the Landau theory has general applicability, we
  show that for the most conducting materials, R=Pr, Nd, both types of
  order can be understood in terms of a nearly-nested spin density wave.
  Furthermore, we argue that in this regime, the charge ordering is
  reliant upon the orthorhombic symmetry of the sample, and therefore
  proportional to the magnitude of the orthorhombic distortion. The
  first order nature of the phase transitions is also explained.  We
  briefly show by example how the theory is readily adapted to modified
  geometries such as nickelate films.
\end{abstract}

\maketitle

The nature of the Mott metal-insulator transition (MIT), driven by
Coulomb repulsion between electrons, is a central subject in condensed
matter physics.  A canonical MIT occurs in the nickelates, perovskites
with the composition RNiO$_3$, where R is a rare earth with nominal
valence R$^{3+}$.  Metallicity in the nickelates correlates with ionic
radius, varying from largest, R=La, which is metallic at all
temperatures, to smallest, R=Lu, which is insulating at all temperatures
; R=Eu has the highest MIT
temperature, T$_{\text{\tiny{MIT}}} =
480$K.\cite{torrance1992systematic} The
nickelates are particularly interesting because they display complex
ordering phenomena in the insulating state.  The insulating ground
states are magnetic, with a large unit cell corresponding to a
periodicity of 4 lattice spacings relative to the ideal cubic
structure, and is usually interpreted in terms of an
``up-up-down-down'' spin configuration.  Remarkably, this complex
ordering seems to be consistent across the entire family.\cite{PhysRevB.64.144417,PhysRevB.57.456,PhysRevB.50.978,garcía1992sudden}  This
magnetism coexists with a form of ``Charge Ordering'' (CO), which modulates
the Ni charge from Ni$^{3+\delta}$ to Ni$^{3-\delta}$ on alternating
sites of a rock-salt type substructure of the cubic perovskite.  In
the more insulating nickelates, R=Eu, Ho, this charge ordering occurs
not only in the ground state but also in an intermediate temperature
insulating phase without magnetism between the high-temperature
metallic phase and the low temperature magnetic one.

It has been postulated that the charge ordering is fundamental, and
should be thought of as separation into spinless Ni$^{4+}$ and spin
$s=1$ Ni$^{2+}$ ions, due to dominant local Hund's rule coupling
$J_H$.\cite{mazin2007charge}  However, no explanation has been given for the particular
complex yet robust magnetic structure in this picture.
Moreover, in the materials PrNiO$_3$ and NdNiO$_3$, for which the MIT
occurs at relative low temperature, the ionic description does not
seem so natural.  In addition, in these materials, charge and spin order occur
simultaneously, bringing the primacy of the former into question.  


In this paper, we introduce a phenomenological Landau theory,
motivated by the particular structure of the e$_g$ bands of the
nickelates near the Fermi energy.  It is particularly relevant to the
more itinerant materials such as PrNiO$_3$ and NdNiO$_3$, where charge
and spin ordering are coincident, and the Landau theory has a natural
microscopic origin in a spin density wave (SDW) picture.  The SDW picture
and Landau theory yield a number of direct insights into the ordering
in bulk samples.  First, the observed unusual magnetic periodicity is
simply explained due to approximate {\sl Fermi surface nesting} of the
$e_g$ bands.  Second, we argue that the observed charge ordering is
naturally induced in this picture {\sl as a secondary order
  parameter}.  Third, the magnitude of the induced charge order is
{\sl proportional to the degree of orthorhombicity} in the material, a
correlation which is indeed observed in experiments, though the
causative relationship does not appear to have been identified
previously.  

The Landau theory can also be easily adapted to other situations.  As
one such application, we derive the a general phase
diagram describing the passage from ``weak'' (SDW-like) Mott
insulators to ``strong'' Mott insulators, in which charge ordering is
indeed dominant, which is  in agreement with that observed in the bulk nickelates, and
we explain the first order nature of the transitions observed there.  
We also show how to incorporate the effects of strain and interfaces
in films and multilayers, and predict that changes of the magnetic
ordering wavevector and exotic ``multiple-q'' states may be thus induced.

{\em Microscopic considerations:} It is helpful to consider as a
semi-microscopic model framework a tight-binding description of the
\eg\ bands,
\begin{equation}
  \label{eq:1}
  H_{tb} = -\sum_{ij} t_{ij}^{ab} c_{ia\sigma}^\dagger c_{jb\sigma}^{\vphantom\dagger},
\end{equation}
where $i,j$ are site indices, $a,b=1,2$ are orbital indices denoting the
two \eg\ states with the symmetry of $2z^2-x^2-y^2$ and $x^2-y^2$
orbitals (these should properly be considered hybridized combinations --
Wannier functions -- of Ni d states and the neighboring O p states, as
RNiO$_3$ is a charge transfer insulator), and
$\sigma=\uparrow,\downarrow$ is the spin index.  Sums over orbital and
spin indices are implied.  The dominant hopping processes are expected
to be those with $\sigma$-type bonding, which can occur with amplitude
$t$ and $t'$ when $i,j$ are first and second neighbor sites,
respectively, of the ideal cubic Ni sublattice.  Specifically, then
$t_{i,i\pm\hat{\mu}}^{ab} = t \phi_\mu^a \phi_\mu^b$ and
$t_{i,i\pm\hat{\mu}\pm\hat{\nu}}^{ab} = t' (\phi_\mu^a\phi_\nu^b +
\phi_\mu^b \phi_\nu^a)$, where $\phi_x =
(-\tfrac{1}{2},\tfrac{\sqrt{3}}{2})$, $\phi_y =
(-\tfrac{1}{2},-\tfrac{\sqrt{3}}{2})$, $\phi_z = (1,0)$ are the orbital
wavefunctions for the $2x^2-y^2-z^2$, $2y^2-x^2-z^2$, and $2z^2-x^2-y^2$
$\sigma$-bonding orbitals along the three axes, in the chosen basis,
respectively.

\begin{figure}
  \centering
\includegraphics[width=3.0in]{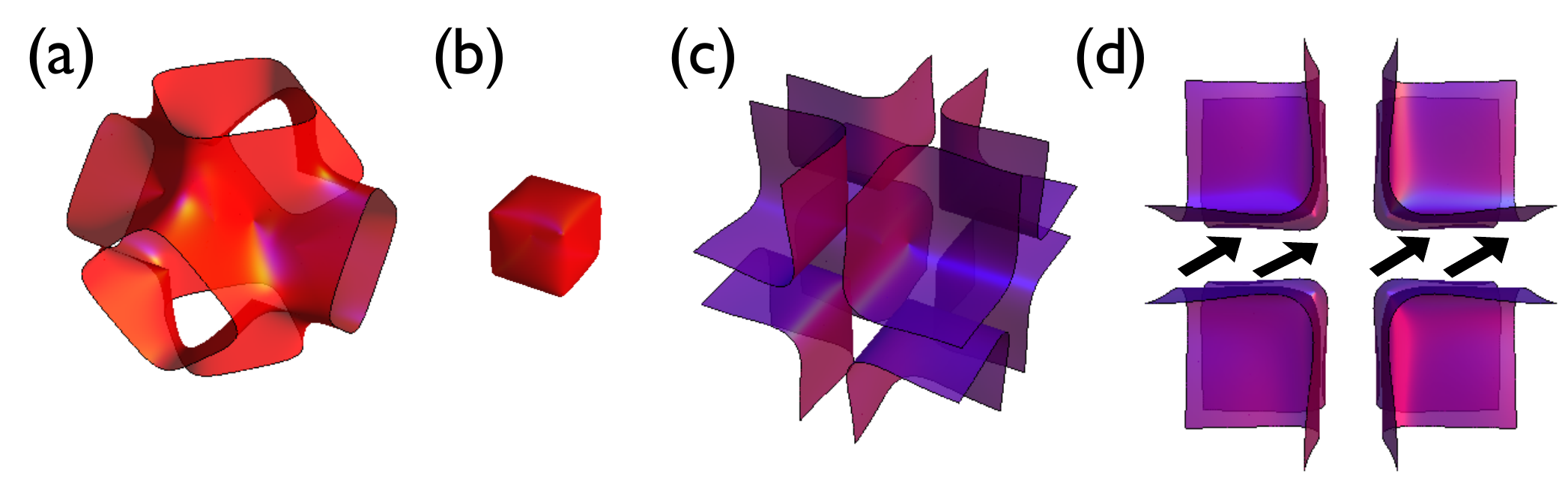}
\caption{Fermi surfaces for the tight-binding model.  In  (a) and (b),
  we show the conduction and valence band Fermi surfaces, respectively,
  for $t'/t = 0.1$.   For larger $t'/t$, the conduction band Fermi
  surfaces become large and hole-like, as shown in (c) and (d) for
  $t'/t=0.3$.  The approximate nesting in the latter case is indicated
  schematically in (d).   }
  \label{fig:FS}
\end{figure}

\begin{figure}
  \centering
\includegraphics[width=3.5in]{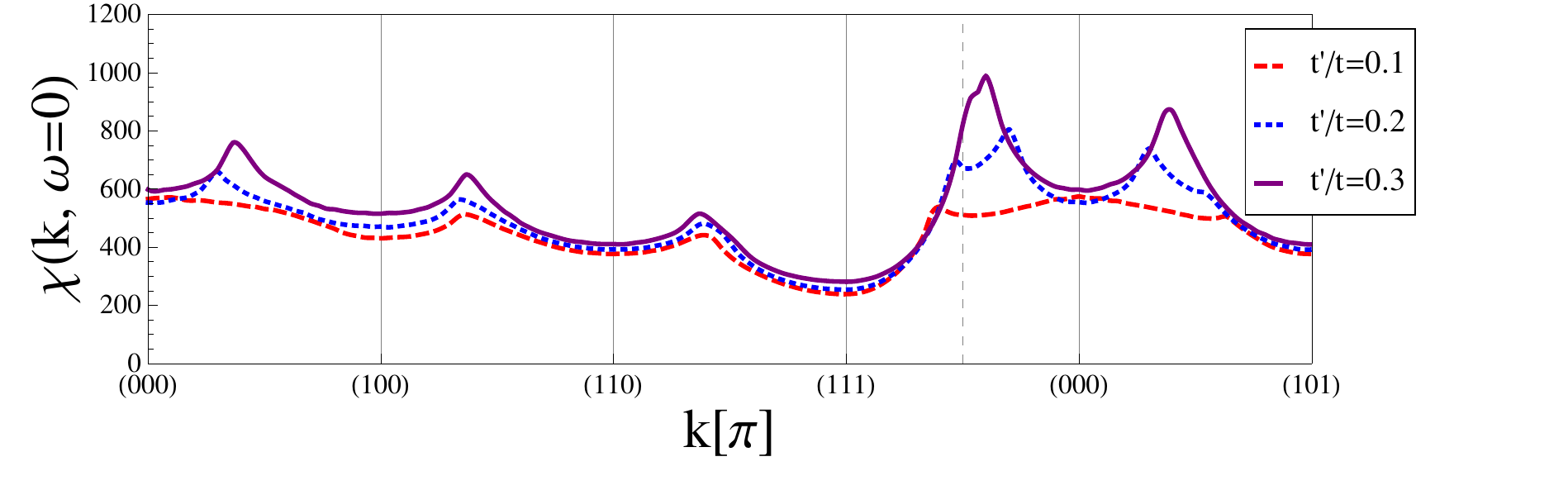}
\caption{Zero frequency spin susceptibility for the tight-binding
  Hamiltonian for $t'/t=0.1,0.2,0.3$, as a function of momentum ${\bf
    k}$ in the cubic Brillouin zone.  Note that for the best nested
  situation, $t'/t=0.3$, the susceptibility is sharply peaked close to
  the wavevector $2\pi(\frac{1}{4},\frac{1}{4},\frac{1}{4})$. }
  \label{fig:chi}
\end{figure}

The above tight-binding model agrees with the \eg\ bands obtained from
LDA calculations,\cite{Hamada19931157} and the Fermi surface measured
recently in photoemission on a LaNiO$_3$ film.\cite{PhysRevB.79.115122}
It also explains well resistivity, Hall effect, and thermopower
measurements on LaNiO$_3$ films,\cite{son:062114} and approximately
describes the inter-band optical spectral weight at energies below about
1eV.\cite{al10:_optic_lanio} LDA favors $t'/t \approx 0.1$, while
comparison to photoemission is best for $t'/t \approx 0.3$, and the
range $0.1\leq t'/t \leq 0.4$ is consistent with transport.  It is
instructive to view the Fermi surface for this range of values (see
Fig.\ref{fig:FS}).  One observes in the middle of this range that the
Fermi surface is rather flat and approximately composed of ``cubes''
rather than spheres.  Fermi surfaces with large flat sections are
considered approximately ``nested'', and well-known to lead to enhance
susceptibilities at certain wavevectors.  Calculation of the free
electron spin susceptibility (see Fig.~\ref{fig:chi}) indeed shows an
enhancement, peaked about wavevectors $\langle k k k\rangle$, with $k$
close to $\frac{1}{4} (\times 2\pi)$ (for $t'/t=0.3$ we find e.g. $k
\approx 0.4\pi$).  With interactions included, the random phase
approximation, or Hartree-Fock theory, both thereby show a tendency to
spin-density-wave (SDW) order at wavevectors close to this one.  It is
remarkable that band theory considerations give a simple mechanism for
SDW order at $k=\pi/2$, which corresponds precisely to that observed in
{\sl all} the insulating nickelates.  Interestingly, we have also found
a mechanism for this order in the strong-coupling limit of the
appropriate multi-orbital Hubbard model, which is too involved to report
here.\cite{lee:_two_hubbar} Together, this may explain the robustness of
the magnetic state observed in experiment.  This unusual magnetic order
has not, to our knowledge, been explained before by any theory.

{\em Landau theory:} Rather than proceeding with a microscopic 
theory (Hartree-Fock and other calculations will be reported in a future
publication\cite{lee:_two_hubbar}), we instead pursue the implications of this view using
symmetry-based Landau analysis.  We begin by considering the problem
for an ideal cubic solid, and take into account distortions of the
perovskite structure at a later stage.  For the SDW order, states with
wavevectors along {\sl any} of the $\langle 111\rangle$ axes are
equivalent.  Hence we actually need to include {\sl four} order
parameters, ${\boldsymbol \psi}_{a}$ with wavevectors ${\bm Q}_a$,
given by ${\bm Q}_0 = \frac{\pi}{2}(1,1,1)$, ${\bm Q}_1
=\frac{\pi}{2}(1,-1,-1)$,${\bm Q}_2 = \frac{\pi}{2}(-1,1,-1)$, and
${\bm Q}_3 = \frac{\pi}{2}(-1,-1,1)$.  Physically, the meaning of the
order parameters, which are complex vectors, is that
\begin{equation}
  \label{eq:29}
  \left\langle {\bf S}_i \right\rangle = \sum_a {\rm Re}\, \left[
    {\boldsymbol \psi}_a e^{i {\bm Q}_a \cdot {\bm r}_i} \right].
\end{equation}

It is a straightforward but lengthy exercise to identify all the allowed
terms, up to fourth order in the SDW order parameters, in the Landau
expansion, taking into account the translational and point group
symmetries of the ideal cubic structure.  At quadratic order, one
obtains only the single coefficient
\begin{equation}
  \label{eq:21}
  F_2 = r \sum_a {\boldsymbol \psi}^*_a \cdot {\boldsymbol
    \psi}_a^{\vphantom*} \equiv r \sum_a |{\boldsymbol \psi}_a|^2.
\end{equation}

At quartic order, one finds 13 distinct terms.  We divide them into
the three terms involving products of only one ``flavor'',
\begin{eqnarray}
  \label{eq:27}
  F_4^{(1)} & = &
  u_1\sum_a\left(
    {\boldsymbol\psi}^*_a\cdot{\boldsymbol\psi}^{\vphantom*}_a\right)^2 +
  u_{2}\sum_a
  \left|{\boldsymbol\psi}^{\vphantom*}_a\cdot{\boldsymbol\psi}^{\vphantom*}_a\right|^2
  \nonumber \\
  && + 
  u_3
  \big(\sum_a \left({\boldsymbol\psi}^{\vphantom*}_a\cdot{\boldsymbol\psi}^{\vphantom*}_a\right)^2
  + {\rm h.c.}\big)  ,
\end{eqnarray}
and ten additional terms involving products of two and four distinct
fields.  For brevity, we will not give these explicitly here, as they
will not play a major role in what follows.  

{\em Single-q states:} The full free energy simplifies greatly if we
restrict to ``single-q'' states, consistent with experiment.  Here
only one of the four order parameters is non-zero, and the ten
quartic Landau invariants {\sl not} shown in Eq.~\eqref{eq:27} vanish.  In a single-q state, we denote below the
non-zero order parameter by simply ${\boldsymbol \psi}$, without
a subscript.

Several types of single-q states are possible, dependent upon values
of $u_2$ and $u_3$ in $F_4^{(1)}$ in Eq.~(\ref{eq:27}).  
The second coefficient, $u_2$, distinguishes between collinear and
spiral states.  For $u_2>2|u_3|$, minimum energy states have ${\boldsymbol
  \psi}= \psi ({\bf \hat n}_1 + i {\bf\hat n_2})$, where ${\bf \hat
  n}_1$ and ${\bf\hat n}_2$ are orthogonal unit vectors.  These describe
coplanar spirals, with spins of fixed length in the plane spanned by
${\bf\hat n}_1$ and ${\bf\hat n}_2$.  The phase of $\psi$ is arbitrary,
corresponding to rotations of the spins within this plane.

For $u_2<2|u_3|$, the free energy is minimized by ${\boldsymbol \psi}
= \psi {\bf\hat n}$, where $\psi$ is a complex scalar.  Such
configurations describe, via Eq.~(\ref{eq:29}), collinear SDW states.
In this case, the remaining coefficient, $u_3$, selects preferential
phases of $\psi$.  Writing $\psi = |\psi|e^{i\theta}$, we finally have
\begin{equation}
  \label{eq:32}
  F_4^{(1)} = (u_1 + u_2)|\psi|^4 + 2 u_3|\psi|^4 \cos 4\theta.
\end{equation}
When $u_3<0$, states with $\theta = \tfrac{\pi}{2} n$ are favored, while
for $u_3>0$, states with $\theta = \tfrac{\pi}{2} (n+\tfrac{1}{2})$ are
favored (in both cases $n=0,1,2,3$ describe four spatially translated
states).  Going back to Eq.~(\ref{eq:29}), and using the wavevector
${\bm Q}_0=\tfrac{\pi}{2}(1,1,1)$ and taking ${\bf\hat n}={\bf\hat z}$
for concreteness, we find that these describe spin states with
\begin{equation}
  \label{eq:33}
  \left\langle S^z_i \right\rangle = \left\{ \begin{array}{cc}
      |\psi|\times (+1,0,-1,0 ,\cdots) & \mbox{for $u_3<0$} \\
      +\tfrac{1}{\sqrt{2}}|\psi|\times (+1,-1,-1,+1, \cdots)  & \mbox{for
        $u_3>0$} \end{array}\right. ,
\end{equation}
where the successive terms in parenthesis describe successive spin
expectation values when moving in unit steps along the principle cubic
$x$, $y$, or $z$ axes in real space.

{\sl Charge order:} For the three types of magnetic states found above
(one spiral, two collinear), let us consider the associated charge
order.   We consider charge ordering at the wavevector $(\pi,\pi,\pi)$,
corresponding to the ``rock salt'' ordering observed in experiment.
The order parameter $\Phi$ is introduced via
\begin{equation}
  \label{eq:2}
  \langle n_i\rangle = \overline{n}+ (-1)^{x_i + y_i + z_i}\Phi,
\end{equation}
where $n_i=\sum_{a\sigma}  c_{ia\sigma}^\dagger c_{ia\sigma}^{\vphantom\dagger}$ is the electron number
operator.   $\Phi$ may equally well be regarded as representing the
amplitude of optical phonons representing octahedral breathing modes
at the same wavevector.  Symmetry allows the following terms in the free energy
involving $\Phi$
\begin{equation}
  \label{eq:3}
  F_\Phi = \tilde{r} \Phi^2 + \tilde{u}\Phi^4 - \lambda \Phi \sum_a
   {\rm Re}\, \left[{\boldsymbol \psi}_a\cdot {\boldsymbol \psi}_a\right],
\end{equation}
where we have included the leading linear coupling to the SDW order
parameters.  The crucial thing to note here is that whenever
${\boldsymbol \psi}\cdot {\boldsymbol \psi}$ is real and non-zero, a
non-zero $\Phi$ is necessarily induced in the minimum free energy
state.  In such situations, $\Phi$ is a {\sl secondary order
  parameter}, slaved to the primary SDW one.  Mathematically, if
$\tilde{u}$ can be neglected, $\Phi$ can be readily ``integrated out''
($F_\Phi$ can be minimized with respect
to $\Phi$) to simply renormalize the quartic SDW couplings.

In this case, we can analyze charge order simply in terms of the SDW
states.  In two of the three SDW states discussed above, the spiral and
the collinear state with $\theta=\pi/4$, ${\rm Re}[{\boldsymbol
  \psi}\cdot {\boldsymbol \psi}]=0$, the charge order vanishes.
This is easily understood since in these cases all sites have equivalent
spin states, up to rotations. In the remaining collinear state, with
$\theta=0$, ${\rm Re}[{\boldsymbol
  \psi}\cdot {\boldsymbol \psi}]=|\psi|^2$ and $\Phi \neq 0$.  Again
this is intuitively clear since the sites with zero spin are obviously
distinct from those with non-zero spin.
Two problems arise now in comparison with
experiment. First, we must arbitrarily choose parameters to be in the
third magnetic state in order to obtain the observed charge order.
Second, in experiment, non-zero magnetic moments, of unequal magnitude,
are clearly observed on the two types of ``inequivalent'' sites, while
in this theoretical configuration the magnetic moment vanishes on half
the sites.

{\em Orthorhombicity: } These problems are resolved by taking into
account the distortions of the ideal perovskite structure.  We focus on
the orthorhombic (Pbnm) structure, which obtains for all the nickelates
except LaNiO$_3$, which is rhombohedral, and does not undergo a Mott
transition.  The orthorhombic distortion is present in the metallic
state up to high temperatures (e.g. up to $T=780$K in
PrNiO$_3$\cite{Lacorre1991225}), and is understood to arise from
reduction in the tolerance factor due to the changing rare earth ionic
radius.  The Pbnm space group has only discrete reflection and inversion
operations in its point group.  The structure has a quadrupled unit
cell, which is doubled by a $45^\circ$ rotated, approximately
$\sqrt{2}\times\sqrt{2}$ enlarged supercell in the $a-b$ plane, and a
doubling along the $c$ axis.  One can rewrite the conventional cubic
coordinates $x,y,z$ in terms of standard orthorhombic coordinates ${\sf
  x,y,z}$, according to $x = {\sf x} + {\sf y}-\frac{1}{2}$, $y = - {\sf
  x} + {\sf y}+\frac{1}{2}$, $z = 2 \,{\sf z}$.  Making this
transformation, one finds that the 4 cubic SDW states corresponds to two
orthorhombic SDW wavevectors: $Q^{\rm ortho}_0=Q^{\rm ortho}_3 =
2\pi(0,\tfrac{1}{2},\tfrac{1}{2})$ and $Q^{\rm ortho}_1=Q^{\rm ortho}_2
= 2\pi(\tfrac{1}{2},0,\tfrac{1}{2})$.  The latter is the wavevector
found in experiments on nickelates with R=Sm, Eu, Nd, Pr, and
Ho.\cite{PhysRevB.64.144417,PhysRevB.57.456,PhysRevB.50.978,garcía1992sudden}

The Pbnm space group has considerably lower symmetry than cubic, containing
only inversion, reflection, and 180$^\circ$ screw axes apart from
translations.  As a consequence, the orthorhombic distortion allows
additional terms in the Landau free energy.  A
straightforward analysis shows that two such terms arise at quadratic order:
\begin{eqnarray}
  \label{eq:36}
  F_2^{\rm ortho} & = & r_1 \big( |{\boldsymbol\psi}_0|^2 -
  |{\boldsymbol\psi}_1|^2 - |{\boldsymbol\psi}_2|^2 +
  |{\boldsymbol\psi}_3|^2 \big) \nonumber \\
  && + r_2 \big( {\boldsymbol\psi}_1\cdot
  {\boldsymbol\psi}_1 - {\boldsymbol\psi}_2\cdot {\boldsymbol\psi}_2 +
  {\rm c.c.}\big) .
\end{eqnarray}
We neglect orthorhombic corrections to the quartic terms on the grounds
that they are presumably smaller.

The terms in Eq.~\eqref{eq:36} clearly lead to an energy difference
between the two possible orthorhombic wavevectors.  We focus on the case
$r_1>-|r_2|$, for which ordering in
${\boldsymbol\psi}_1,{\boldsymbol\psi}_2$ is preferred, which
corresponds to the experimental wavevector.  To proceed, we suppose, without loss of generality, that ${\boldsymbol
  \psi}_1={\boldsymbol \psi}$ is the non-zero order parameter.  For the
spiral state, we now consider the cubic energy from
Eq.~(\ref{eq:27}), with $u_2>0$, added to the $r_2$ term.  For
$r_2$ non-zero, the minimum energy configuration is deformed to
${\bf \psi} \propto (1+\delta) {\bf
  \hat n}_1 + i (1-\delta) {\bf \hat n}_2$, with $\delta \propto r_2$.
This corresponds to a deformed spiral in which spins trace out an
ellipse rather than a circle.  In this state, $\Phi\propto {\boldsymbol
  \psi}\cdot {\boldsymbol \psi}\propto r_2$, so charge order is
induced.

For the collinear state, Eq.~\eqref{eq:32} is
modified to
\begin{equation}
  \label{eq:37}
  F(\theta) = F_0 + 2 u_3 |\psi|^4 \cos 4\theta + 2 r_2 |\psi|^2 \cos 2\theta.
\end{equation}
There are two cases to consider here.  For $u_3<0$, both cosines can be
simultaneously minimized.  The minimum in this case occurs at either
$\theta=\tfrac{\pi}{2}n$, with integer $n$ either even or odd if $r_2$
is negative or positive, respectively.  This selects the
$+1,0,-1,0,\cdots$ type ordering in the first line of
Eq.~(\ref{eq:33}).

In the other case, $u_3>0$, the situation is more interesting.  The
minimum at $\theta = \tfrac{\pi}{2} (n+1/2)$ are {\sl unstable} to small
$r_2$ of either sign.  Thus the generic situation in this case, at least
for small $|r_2|$, is to obtain a generic value of $0<\theta<\pi/4$.
(For sufficiently large $r_2$, the minimum $\theta$ will again lock to
the above type of solution).  As a result, one obtains the type of order
observed in experiment.

The conclusion is that in the $Q^{\rm
  ortho}=2\pi(\tfrac{1}{2},0,\tfrac{1}{2})$ state, the SDW is {\sl
  always} accompanied by charge order due to orthorhombicity.
Moreover, if the state is such that moments are present on all sites,
then the charge order is {\sl proportional} to the degree of
orthorhombicity, parametrized here by $r_2$.  This is an important
conclusion of this paper.  Interestingly, if the analysis is repeated
for a {\sl rhombohedral} (R$\overline{\rm 3}$c) crystal, no charge
order is induced in this way.\cite{lee:_two_hubbar}  Hence for a rhombohedral
insulator,\cite{Vassiliou1989208} we predict the occurrence of a
``pure'' SDW state.

\begin{figure}
  \centering
\includegraphics[width=2.7in]{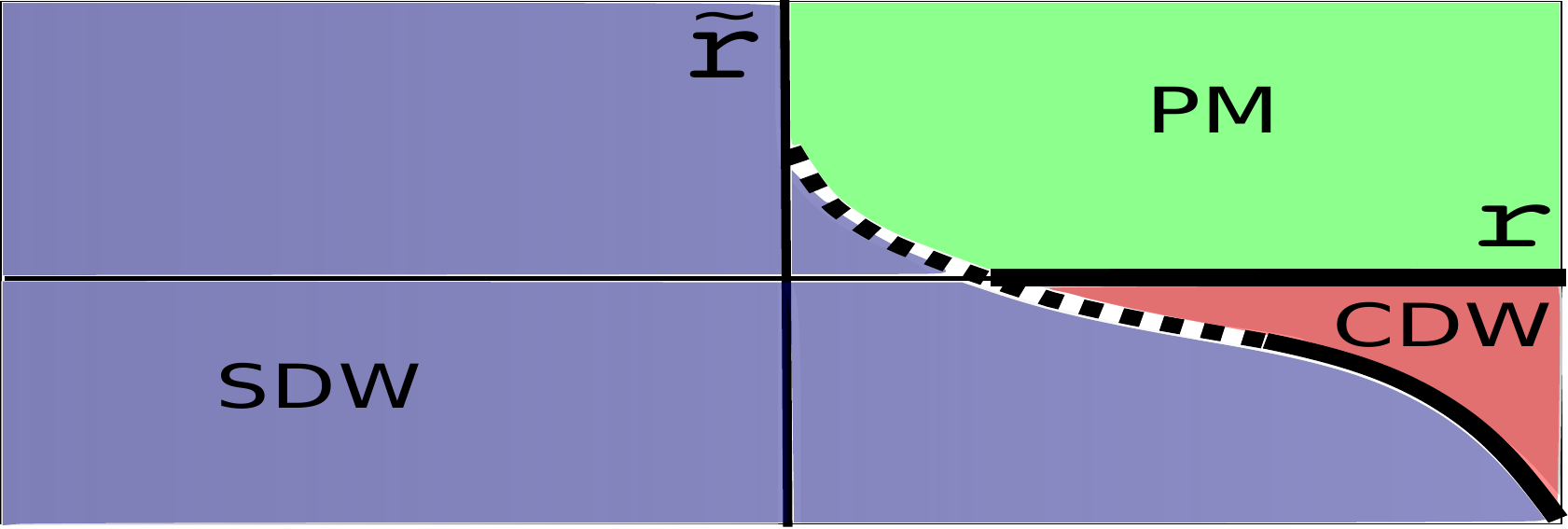}
\caption{ Mean-field phase diagram for the Landau theory of
  Eq.~\eqref{eq:5}, describing the interplay of spin and charge
  ordering. The dashed (solid) lines delineate first (second) order
  transitions within mean field theory. }
  \label{fig:landaupd}
\end{figure}

{\em Order of transitions: }  The three phase transitions -- metal-CO,
metal-SDW, and CO-SDW -- observed in experiment are all first order.
This behavior can be rationalized by various mechanisms.  First consider
a mean-field treatment of the Landau free energy.     For simplicity,
we assume single-q states, and neglect the distinctions between
collinear and spiral spin states, taking ${\boldsymbol \psi}=\psi
{\bf\hat x}$ with $\psi$ real.  Then the full Landau free energy is
\begin{equation}
  \label{eq:5}
  F = r \psi^2 + u \psi^4 - \lambda \Phi \psi^2 + \tilde{r}\Phi^2 + \tilde{u}\Phi^4.
\end{equation}
By minimizing this free energy, one obtains the phase diagram shown in
Fig.~\ref{fig:landaupd}.  For $\lambda=0$, the SDW and CO orders are
tuned independently by $r$ and $\tilde{r}$ respectively.  However,
with non-zero spin-charge coupling $\lambda$, the region near the
origin is modified.  In particular, the consequence is that for
systems close to the multicritical point at which the CO phase
emerges, all the transitions to the SDW phase become first order.
This may explain the discontinuous SDW transitions seen in
experiment. To explain the first order metal-CO transition, we note
that electron-phonon interactions are significant in
nickelates.\cite{PhysRevLett.80.2397} In particular coupling of $\Phi$
to {\sl acoustic phonons} renders the theory of the metal-CO
transition equivalent to the {\sl compressible Ising model}, which is
known to have a first order transition.\cite{PhysRevB.13.2145}

{\em Films: } We conclude with examples of how the Landau theory is adapted to
modified geometries.  Pure strain is relevant to thick
films in which interfaces are not important.  For a tetragonal
substrate, symmetry implies that the strain modifies $r$
(Eq.~\eqref{eq:21}) and $r_1$ (Eq.~\eqref{eq:36}).  If the intrinsic
orthorhombic contribution to $r_1$ is small, we predict that strain
can switch the wavevector from $Q_{1/2}^{\rm ortho}$ to $Q_{0/3}^{\rm
  ortho}$, and at the same time ``turn off'' the charge ordering.  A
more severe effect occurs at an interface, due to the lack of
translational symmetry normal it.  This leads to symmetry-allowed terms
which {\sl mix} the SDW states
with wavevectors $Q_1$ and $Q_2$ (or $Q_0$ and $Q_3$),
so that ``multiple-q'' ordering appears near the interface.

\acknowledgements

We are grateful to Susanne Stemmer, Jim Allen, Dan Ouellette, and Junwoo
Son for discussions and experimental inspiration.  This work was
supported by the NSF through grants PHY05-51164 and DMR-0804564, and
the Army Research Office through MURI grant No. W911-NF-09-1-0398.

\bibliography{landaun}

\end{document}